\documentclass[cameraready]{Interspeech}

\title{Deep Learning Based Relative Transfer Matrix Estimation for Multiple Sources and Multiple Microphones}

\author[affiliation={1}, orcid=0009-0002-6357-6819, correspondingauthor]{Oshan A.~B.}{Yalegama}
\author[affiliation={1,2}, orcid=0000-0002-5864-4154
]{Wageesha N.}{Manamperi}

\address{
    $^1$ Department of Electronic and Telecommunication Engineering, University of Moratuwa, Sri Lanka \\
    $^2$ School of Engineering, The Australian National University, Australia   
}

\email{yalegamammoab.20@uom.lk, wageesham@uom.lk}


\usepackage{comment}
\usepackage{multirow}

\begin{document}

\maketitle

\begin{abstract}
The Relative Transfer Matrix (ReTM), recently introduced as a generalization of the relative transfer function for multiple receivers and sources, shows promising performance when applied to speech enhancement in noisy environments. Estimating the ReTM of sound sources by exploiting the covariance matrices of multichannel recordings is highly beneficial for practical applications and, to date, remains the only proposed approach. This paper investigates deep learning-based ReTM estimation. We propose three novel supervised learning frameworks using time and short-time frequency transform domain convolutional networks, and a Long Short-Term Memory-based recurrent neural network. Experimental results demonstrate that the proposed models achieve more accurate estimation of the ReTM using five objective metrics compared to the covariance-based method. We also show the effectiveness of the proposed frameworks for speech enhancement, achieving performance on par with the baseline method.
\end{abstract}

\section{Introduction}
Blind estimation of the relative transfer function (ReTF) without requiring positional knowledge of receivers or sound sources is highly attractive in practical applications such as robot audition \cite{nakadai2000active}, drone audition \cite{RN223}, teleconferencing \cite{araki2008speaker}, and hearing aids \cite{marquardt2016incorporating}, which involve tasks such as sound source localization, speech enhancement, speaker separation, and acoustic echo cancellation \cite{gannot2001signal,talmon2009relative,warsitz2008speech,wageeshadrtf}.
However, these applications typically involve multiple simultaneously active sound sources, where the assumption of W-disjoint orthogonality \cite{yilmaz2004blind} required for ReTF estimation does not hold.
To overcome this limitation, the relative transfer matrix (ReTM) \cite{abhayapala2023generalizing} has been introduced as a generalization of the ReTF for acoustic environments with multiple simultaneously active sound sources and multiple microphones, providing an essential component in audio signal processing techniques and applications.
In this paper, we propose machine learning-based methods for ReTM estimation from microphone recordings.

Many approaches to estimate ReTF for a single active sound source have been developed \cite{gannot2001signal,talmon2009relative,warsitz2008speech,serizel2014low,varzandeh2017iterative,markovich2009multichannel,middelberg2023relative,warsitz2007blind,markovich2018performance,shalvi1996system,cohen2004relative,brendel2022manifold,sofer2021robust,yang2021supervised,yang2021learning,birnie2021noise,wang2018mask,levi2025peerrtf}. Among them, covariance-based methods have gained attention due to their practical feasibility as well as simplicity \cite{serizel2014low,varzandeh2017iterative,markovich2009multichannel,middelberg2023relative,warsitz2007blind,markovich2018performance}, whereas, machine learning-based approaches offer superior performance at an increased computational cost \cite{brendel2022manifold,sofer2021robust,yang2021supervised,yang2021learning,birnie2021noise,wang2018mask,levi2025peerrtf}.

In \cite{abhayapala2023generalizing}, the concept of ReTM was introduced to relate the signal received between two sets of microphone groups with respect to all active sources present in a room. Similar to the ReTF \cite{talmon2009relative}, the ReTM is independent of source signals but dependent on the spatial location of the sources and the environment \cite{abhayapala2023generalizing}.
Recently, various studies have been proposed to utilize ReTM for speech enhancement and speaker separation in multi-source noisy reverberant environments \cite{wnmanamperiReTMdDrone, kumar2024speech, retmcovadd, wnmanamperiReTM2ReTF, manamperi2025multiple}, which have shown significant performance improvement. Traditional ReTM estimation approach exploits statistical relationships between two microphone groups using covariance matrices \cite{abhayapala2023generalizing}. In contrast to the baseline method that learns a spatial mapping between receiver groups, alternative approaches remain largely unexplored. 

In this paper, we introduce three fully supervised machine learning frameworks for ReTM estimation. We use both time and short-time Fourier transform (STFT) domains, deep learning approaches. All three models assume a stationary environment that all sound sources are unmoving. The contributions of this work are as follows: 1) We propose: i) the STFT Convolutional Network (SCoNet) that uses depthwise convolution;
ii) the Convolutional Filter and Summation Network (FuSNet) that applies a set of individual convolutional filters; and iii) the Long Short-Term Memory (LSTM)-based Autoencoder Network (LAeNet) that uses a shared bidirectional-LSTM followed by a feedforward network. 2) We evaluate ReTM estimation accuracy with respect to the covariance-based approach using five qualitative metrics, signal-to-distortion ratio (SDR), mean square error (MSE), log spectral distortion (LSD), average phase distortion (APD), and relative spectrum error (RSE). 3) We demonstrate the merits of the proposed models in speech denoising. 
The simulation results are comparable to the baseline method overall, while the STFT domain models consistently outperform it.

\section{Problem formulation}
\label{sec:prob_form}

Consider a reverberant environment with concurrently active $\mathcal{L}$ sound sources.
In the short time Fourier transform (STFT) domain, we denote $S_\ell(f,t) $, $\ell = {1,\cdots,\mathcal{L}}$ as the source signals.
Let there be $Q$ arbitrary distributed microphones in the room. We divide them to two groups of microphones, $\{A\}$ and $\{B\}$ with $Q_A$ and $Q_B$ microphones, respectively ($Q = Q_A + Q_B$).
We denote $\mathbf{M}_A(f,t) $ and $\mathbf{M}_B(f,t) $  as the vector of received signals at microphone groups A and B, respectively.
Then the received signals at each microphone group in matrix form as
%
\begin{equation}\label{eqn:M_A}
    \mathbf{M}_A(f,t) = \mathbf{H}_A(f)\mathbf{S}(f,t), 
\end{equation}
\begin{equation}\nonumber
    \mathbf{M}_B(f,t) = \mathbf{H}_B(f)\mathbf{S}(f,t),
\end{equation}
where 
$\mathbf{S}(f,t) = [S_1(f,t), \ldots, S_{\mathcal{L}}(f,t)]^T$, and $\{\cdot\}^T$ is  the matrix transpose.
Here, $\mathbf{H}_A(f) \in \mathbb{C}^{Q_A \times \mathcal{L}}$ and $\mathbf{H}_B(f) \in \mathbb{C}^{Q_B \times \mathcal{L}}$ are the matrices with elements defined by the acoustic transfer functions. Note that we consider thermal microphone noise to be negligible.

The ReTM, $\bm{\mathcal{R}}_{AB}(f)$, is defined as in \cite{abhayapala2023generalizing} \begin{equation}\label{eqn:retm_hahb}
    \bm{\mathcal{R}}_{AB}(f) \triangleq \mathbf{H}_A(f) \mathbf{H}_B(f)^\dagger,
\end{equation}
where $(\cdot)^\dagger$ is Moore-Penrose inverse, assuming the validity, i.e., $Q_B \geq \mathcal{L}$. Thus, we can relate the received signal at group $\{A\}$ and $\{B\}$ using
%
\begin{equation} 
\label{retm_freq}
     \mathbf{M}_A(f,t)  = \bm{\mathcal{R}}_{AB}(f) \mathbf{M}_B(f,t).
\end{equation}

In time domain Equation~\ref{retm_freq} is given by
\begin{equation}
\label{retm_time}
m_A^j(t) = \sum_{i=1}^{Q_B} r_{AB}^{ji}(t) \ast m_B^i(t), \quad j = {1,\cdots,Q_A},
\end{equation}
where $m_A^j$, and $m_B^i$ denote the signals of the $j^{\text{th}}$ microphone in group $\{A\}$, and the $i^{\text{th}}$ microphone in group $\{B\}$, respectively, $r_{AB}^{ji}$ denotes the inverse Fourier transform of the $\{j,i\}^{\text{th}}$ element of $\boldsymbol{\mathcal{R}}_{AB}(f)$.

The aim of this paper is to exploit the spatial properties of sound sources by modeling the ReTM with machine learning algorithms in the time domain and STFT domain as $f_B(\cdot)$, and $F_B(\cdot)$, respectively, such that 
\begin{equation}
\label{function_freq}
\bm{M}_A(f,t) = F_B(\bm{M}_B(f,t)),
\end{equation}
\begin{equation}
\label{function_time}
\boldsymbol{m}_A(t) = f_B(\boldsymbol{m}_B(t)).
\end{equation}
The next section proposes machine learning approaches for the modeling of $f_B(\cdot)$ and $F_B(\cdot)$.

\vspace{-0.3cm}
\section{Proposed ReTM estimation models}\label{models}

In this section, we present two convolutional networks and one Long Short-Term Memory (LSTM)-based recurrent neural network for the modeling of the ReTM. 

\vspace{-0.2cm}
\subsection{STFT convolutional network (SCoNet)}

\begin{figure}[t]
  \centering
  \centerline{\includegraphics[width=\columnwidth]{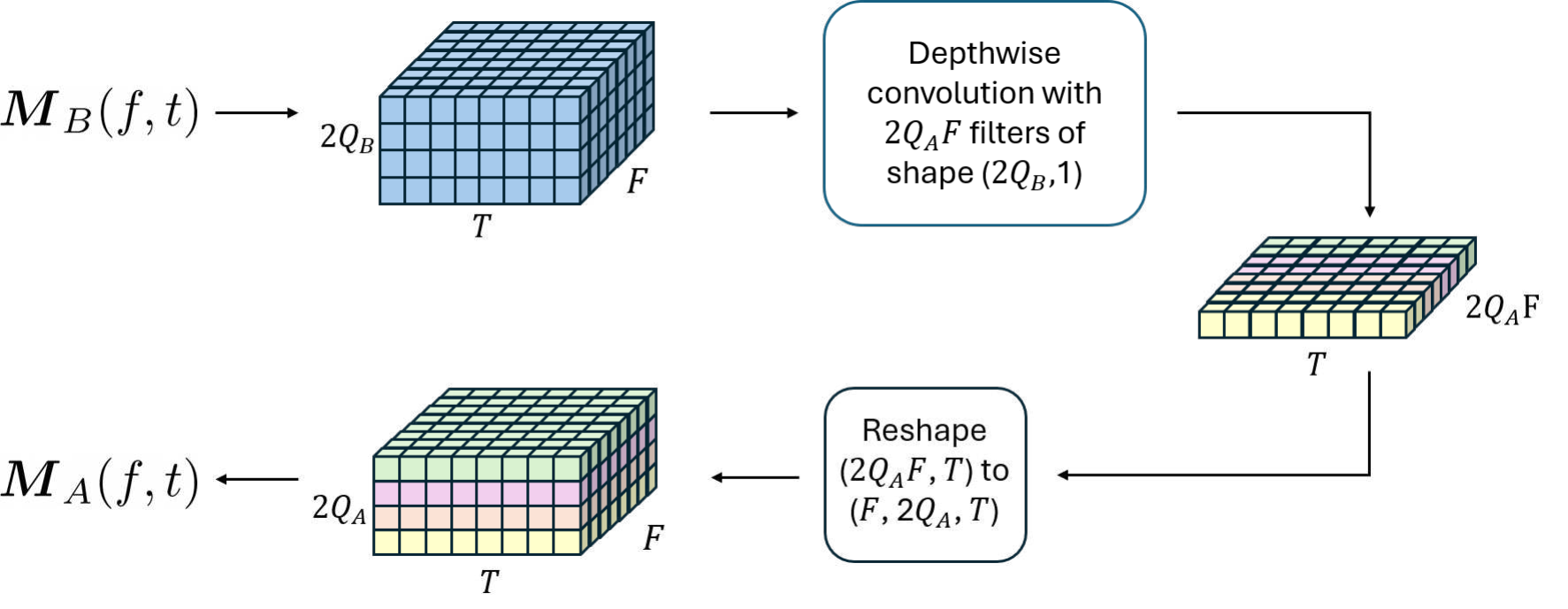}}
  \caption{Model architecture of SCoNet, where $F$ and $T$ are the number of frequency bins and time frames, respectively. }
  \label{fig:cnnstft}
\end{figure}

We model $F_B(\cdot)$ using SCoNet, which applies depthwise convolution operations in the STFT domain, as shown in Figure~\ref{fig:cnnstft}. 
We input $\boldsymbol{M}_B(f,t)$, STFT of microphone signals in group $\{B\}$, and train to estimate the microphone signals in group $\{A\}$. SCoNet stacks the real and imaginary parts along the channel dimension and performs two-dimensional depthwise convolution across the time and channel axes, allowing the model to learn distinct parameter sets for each frequency component of the relative transfer matrix. The channel-wise outputs of group $\{A\}$ are then stacked to form the STFT representation of $M_A(f,t)$.

\subsection{Convolutional filter and summation network (FuSNet)}

\begin{figure*}[htb]
    \begin{minipage}[b]{0.48\linewidth}
        \centering
        \includegraphics[width=\linewidth]{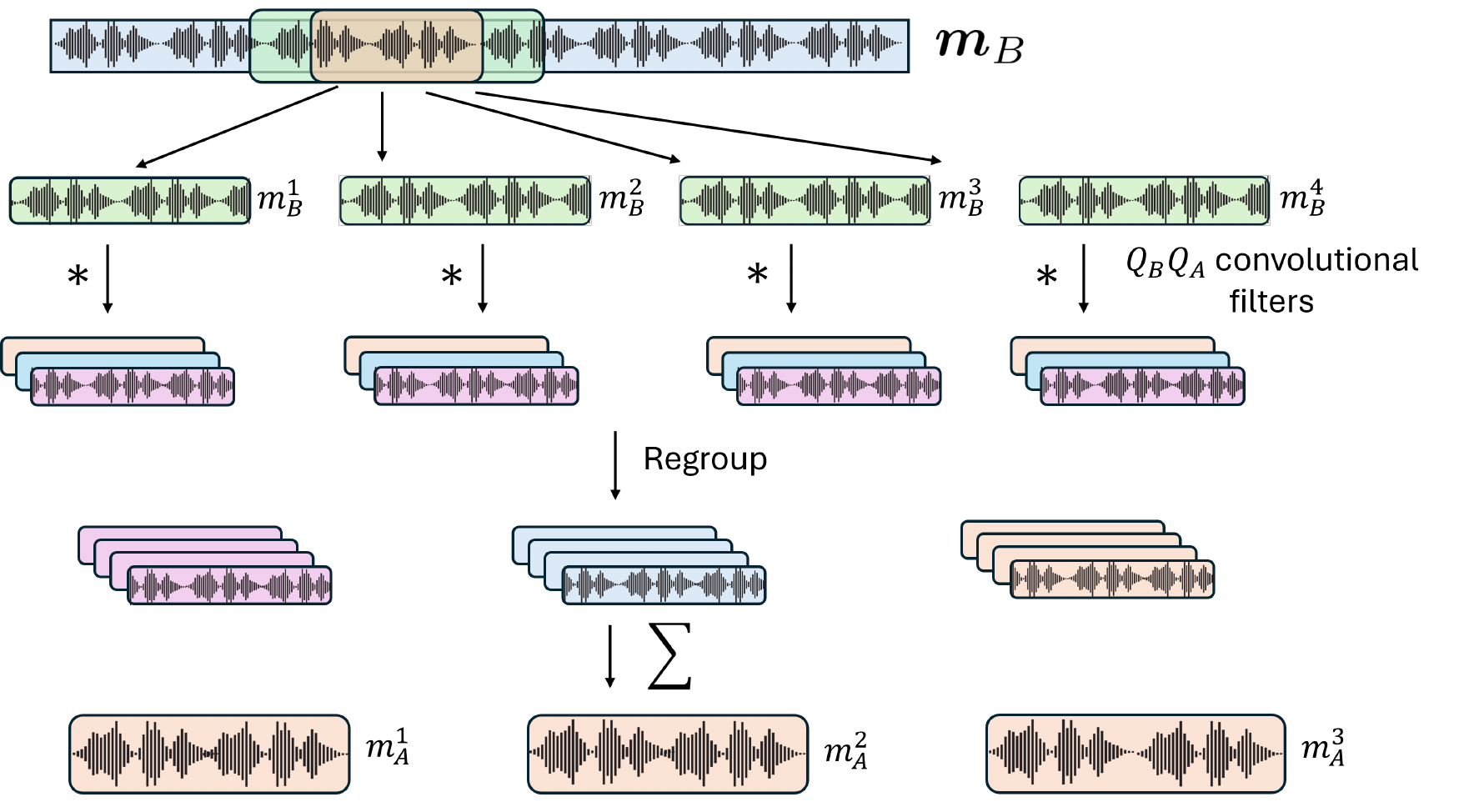}
        \caption{Model architecture of FuSNet for the case of $Q_B=4$ and $Q_A =3$.}
        \label{fig:fasnet}
    \end{minipage}
    \hfill
    \begin{minipage}[b]{0.48\linewidth}
        \centering
        \includegraphics[width=\linewidth]{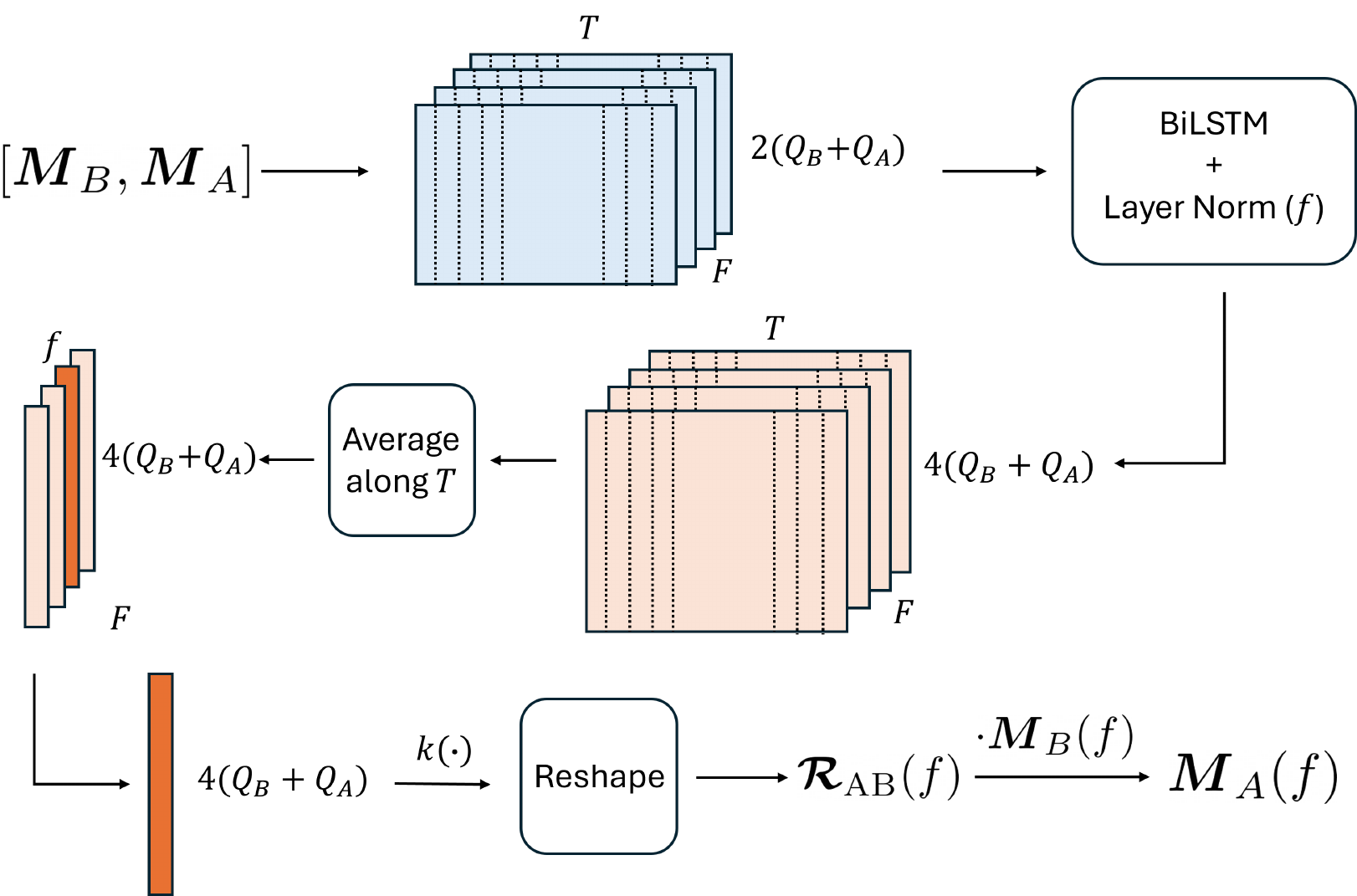}
        \caption{Model architecture of LAeNet with a shared  BiLSTM. }
        \label{fig:lstm}
    \end{minipage}
\end{figure*}

Motivated by the effectiveness of convolutional neural networks (CNNs) in modeling the ReTF~\cite{birnie2021noise}, we propose FuSNet, which parameterizes $f_B(\cdot)$ using $Q_A \times Q_B$ learnable one-dimensional convolutional filters, whose weights correspond to $r_{AB}^{ji}(t)$ in Equation~\ref{retm_time}. Figure~\ref{fig:fasnet} shows the FuSNet architecture. We input non-overlapping segments of length $L$, and a context window of size $3L$ from the time frames of microphone signals at group $\{B\}$. The context window is chosen such that the convolution output matches the original segment length, setting the filter size to $L$. Note that the window length should be longer than the room’s reverberation time to satisfy the multiplicative transfer function \cite{birnie2021noise, avargel2007multiplicative}. The convolution outputs are regrouped and summed to obtain the time domain microphone signals at group $\{A\}$. 

\subsection{LSTM based autoencoder network (LAeNet)}

We propose LAeNet, an LSTM-based autoencoder for modeling of $F_B(\cdot)$, as shown in Figure~\ref{fig:lstm}. This design exploits LSTMs’ ability to capture narrow-band spatial information~\cite{yang2023mcnet}. In LAeNet, each frequency bin of the concatenated STFTs $\boldsymbol{M}_A(f,t)$ and $\boldsymbol{M}_B(f,t)$ is independently processed by a bidirectional-LSTM (BiLSTM) layer with shared weights. The resulting temporal sequences are passed through layer normalization to stabilize and accelerate training, then averaged over time to yield feature vectors of size $4(Q_B+Q_A)$, where the increase in dimensionality stems from the BiLSTM. Each feature vector is subsequently mapped by a fully connected network $k(\cdot)$ to estimate the ReTM coefficients, which are finally used to estimate the group $\{A\}$ microphone signals $\boldsymbol{M}_A(f,t)$ as in Equation~\ref{retm_freq}.


\subsection{Loss function and training}

To improve the stability and performance of the proposed models, we train using a weighted sum of the negative signal-to-distortion ratio (SDR) computed in the time domain, and the relative spectrum error (RSE) measured in the STFT domain as:
\begin{equation}
L = -\alpha L_{SDR}
(\boldsymbol{m}_A, \hat{\boldsymbol{m}}_A) + \beta L_{RSE}(\boldsymbol{M}_A, \hat{\boldsymbol{M}}_A),
\end{equation}
where $\hat{\boldsymbol{m}}_A$, and $\hat{\boldsymbol{M}}_A$ denote the estimated microphone signals at group $\{A\}$ in the time and STFT domains, respectively. This joint optimization enforces consistency across both representations and improves robustness to noise and artifacts that may manifest differently in the time and frequency domains. Here, $\alpha$ and $\beta$ are weight factors for
\begin{equation}
    L_{SDR} = \frac{1}{T} \sum_{t=1}^{T} 10 \log_{10} \left( \frac{|m_A(t)|_2^2}{|m_A(t)- \hat{m}_A(t)|_2^2} \right)
\end{equation}
and 
\begin{equation}
    L_{RSE} = \frac{1}{FT}
\sum_{t=1}^{T}
\sum_{f=1}^{F}10 \log_{10} \left( \frac{|\hat{M}_A(f,t) - M_A(f,t)|^2}{|M_A(f,t)|^2} \right),
\end{equation}
respectively. For training, we employ the Adam optimizer \cite{kingma2014adam} with a learning rate that is adaptively reduced over epochs based on the validation set performance.

\section{Experiments}\label{results}


\subsection{Experimental methodology}

We utilize an open-source toolbox \cite{rirgen} to model the room impulse response from the sound sources to irregularly distributed microphones in a  $6 \times 7 \times 3$ m rectangular room ($T_{60} = 500$ ms). We assign $Q = 7$ with $Q_A = 3$, and $Q_B = 4$ number of receivers to group $\{A\}$ and $\{B\}$, respectively. We consider four scenarios. \textbf{A1}: two white Gaussian noise (WGN) sources, \textbf{A2}: two noise sources (air conditioner noise, music), and \textbf{B}: three sources (speech, air conditioner noise, music). Scenario \textbf{C} uses a total of $Q =12$ microphones with $Q_A = 5$, and $Q_B = 7$, drives with the same set of sources as \textbf{B}. The received signals are added with $40$ dB SNR of WGN at each microphone and down-sampled to $16$ kHz. For Scenario A1, the training, validation, and test recordings are 3 minutes, 1 minute, and 1 minute, respectively. For all other scenarios, 50 s and 10 s recordings are used for training and testing, respectively.

For the model hyperparameter configuration, FuSNet is designed with a window and context size of 8192 samples. For SCoNet and LAeNet, the input recordings are converted to STFT frames using a Hann window of size 8192 with 50\% overlap. The window length is selected based on the room’s reverberation time to ensure that it is sufficiently long for the multiplicative transfer function assumption to hold \cite{avargel2007multiplicative}. The weight factors $\alpha$ and $\beta$ are set to 1 and 10, respectively. 

For a comprehensive evaluation of the proposed models, we adopt the covariance-based approach~\cite{abhayapala2023generalizing} as the baseline. To fairly assess performance across all scenarios, we use five quantitative metrics: 
(i) SDR, (ii) mean square error (MSE), defined as 
\[
\text{MSE} = 10\log_{10}\left(\frac{1}{T}\sum_{t=1}^{T}|m_A(t) - \hat{m}_A(t)|^2\right),
\]
(iii) log spectral distortion (LSD), defined as
\[
\text{LSD} = \sqrt{\frac{1}{N} \sum_{\eta=1}^{N} \left( 10 \log_{10} {|M_A(f_\eta, t)|^2 /|\hat{M}_A(f_\eta, t)|^2} \right),^{2}}
\]
(iv) average phase distortion (APD), defined as
\[
\text{APD} = \frac{1}{N} \sum_{\eta=1}^{N} \left| \angle M_A(f_\eta, t) - \angle \hat{M}_A(f_\eta, t) \right|,
\] 
and (v) RSE.

\subsection{Results and discussion}\label{results_and_discussion}

\begin{table*}[ht]
\centering
\caption{ReTM estimation accuracy for various scenarios (Channel 1/Average).}
\label{tab:metrics}
\begin{tabular}{|c|c|c|c|c|c|c|}
\hline
Scenario & Method & SDR (dB)↑ & MSE (dB)↓ & LSD (dB)↓ & APD (rad)↓ & RSE (dB)↓\\
\hline
\multirow{4}{*}{\textbf{A1}} 
& Baseline      &  $26.56 /26.1$   & $-43.84 / -43.56$ & $1.21 / 1.19$ &  $0.1 / 0.1$ & $-24.76 / -24.10$\\
& SCoNet  & $24.76 / 24.24$  & $-40.88 / -40.54$ &  $0.89 / 0.96$ & $0.13 / 0.14$ &  $-25.62 / -24.89$\\
& FuSNet &  $\bm{29.13 / 28.46}$  &  $\bm{-59.25 / -58.75}$ &  $\bm{0.5 / 0.49}$ &  $\bm{0.08 / 0.09}$ &  $\bm{-48.61 / -48.4}$ \\
& LAeNet  & $25.12 / 24.09$  &  $-55.24 / -54.38 $ & $1.03 / 1.13$ & $0.12 / 0.13$ & $-20.25 / -19.74$\\
\hline
\multirow{4}{*}{\textbf{A2}} 
& Baseline      & $21.2 / 20.5$    & $-36.25 / -36.57$ & $2.02 / 2.03$ & $0.21 / 0.21$ &  $-25.28 / 24.65$\\
& SCoNet  &  $26.26 / 26.59 $ & $-43.08 / -42.47$ &  $\bm{1.22 / 1.23}$ &  $\bm{0.18 / 0.19}$ & $-24.61 / -24.06$\\
& FuSNet &  $\bm{30.96 / 29.62}$  &  $\bm{-55.2 / -54.94}$ & $2.47 / 2.59$ & $0.23 / 0.2$ &  $\bm{-39.81 / -33.78}$ \\
& LAeNet  &  $28.07 / 26.21$  &  $-52.32 / -51.53$ &  $1.68 / 1.70$ & $0.22 / 0.21$ & $-18.17 / -17.67$\\
\hline
\multirow{4}{*}{\textbf{B}} 
& Baseline     & $16.1 / 16.29$   & $-36.65 / -37.49$ & $2.48 / 2.5$ & $0.23 / 0.24$ & $-17.39 / -16.85$\\
& SCoNet  &  $\bm{22.42 / 21.96}$  & $-42.78 / -42.98$ &  $\bm{1.53 / 1.54}$ &  $\bm{0.19 / 0.19}$ &  $\bm{-20.76 / -20.33}$\\
& FuSNet &  $22.51 / 21.91 $ &  $\bm{-45.12 / -45.17}$ & $4.89 / 4.96$ & $0.71 / 0.72$ &   $-19.60 / -19.16$\\
& LAeNet  & $22.36 / 21.88$  &  $-45.01$ / $\bm{-45.18}$ &  $1.78 / 1.88$ &  $0.2 / 0.21$ & $-17.58 / -17.21$\\
\hline
\multirow{4}{*}{\textbf{C}} 
& Baseline      & $23.05 / 23.25$  & $-43.6 / -43.35$ &  $\bm{1.03 / 1.09}$ & $ 0.15 / 0.15$ &  $-34.03 / -33.22$\\
& SCoNet  &  $29.59 / 27.85$  & $-44.12 / -43.27$ &  $1.06 / 1.17$ & $\bm{0.14}$ / $0.16$ & $-28.63 / -27.43$\\
& FuSNet &  $\bm{40.33 / 38.88}$  &  $\bm{-62.89 / -61.98}$ & $1.95 / 2.10$ & $0.15 / 0.16$ &  $\bm{-34.07 / -33.27}$\\
& LAeNet  &  $29.35 / 28.57 $ &  ${-52 / -51.77}$   & $1.39 / 1.49$ &  $\bm{0.14 / 0.15}$ & $-20.15 / -19.97$\\
\hline
\end{tabular}
\end{table*}

The ReTM estimation accuracy results are given in Table~\ref{tab:metrics}. From the results in scenario \textbf{A1}, we observe that FuSNet achieves the highest performance across both time and frequency domain metrics. This improvement of FuSNet can be attributed to the use of WGN sources that provide a uniform spectral distribution and thereby facilitate more accurate frequency domain estimation. The other proposed methods, SCoNet and LAeNet, also achieve high estimation accuracy, performing comparably to the baseline method.

In scenario \textbf{A2}, FuSNet again achieves the highest overall performance across most evaluation metrics, with noticeable degradation in LSD and APD. The key difference between scenarios \textbf{A1} and \textbf{A2} lies in the replacement of WGN with specific noise sources, which exhibit non-uniform frequency distributions. This indicates that FuSNet performs best with uniform spectra but degrades under non-uniform conditions. Moreover, SCoNet yields performance on par with LAeNet and surpasses the state-of-the-art method, excluding the RSE score, where the baseline approach shows the second-best performance.

From scenarios \textbf{B} and \textbf{C}, we observe that all four methods exhibit improved ReTM estimation accuracy as the number of microphones in both groups increases, consistently across all five evaluation metrics. 
The traditional method exhibits significantly reduced accuracy, and is seen to be the worst performing method among all methods, except for the LSD and RSE measures. Compared to the baseline, both SCoNet and FuSNet, show substantially superior performance. SCoNet performs best in scenario \textbf{B} with an MSE score lower than FuSNet. Whereas, FuSNet achieves the highest performance in scenario \textbf{C}, excluding the LSD and APD measures.

Overall, it can be verified that although the baseline method achieves satisfactory results, it is clearly outperformed by the proposed deep learning-based models, particularly in the time domain measures, e.g., MSE and SDR.

\begin{table}[]
\centering
\caption{Computational complexity of each proposed model.}
\label{table:parameters}
\begin{tabular}{c|cc|cc}
\hline
\multirow{2}{*}{Method} 

  & \multicolumn{2}{c|}{ $QA=3, Q_B=4$} 
  & \multicolumn{2}{c}{ $Q_A=5, Q_B=7$} \\
\cline{2-5}
  & \multicolumn{1}{c}{Param.} 
  & \multicolumn{1}{c|}{Latency} 
  & \multicolumn{1}{c}{Param.} 
  & \multicolumn{1}{c}{Latency} \\
\hline
SCoNet   &  $196.7$k & 6.77ms &  $573.6$k & 6.61ms\\
FuSNet  & $98.3$k & 1.07ms  & $286.8$k & 2.41ms\\
LAeNet   & $263.5$k & 1.80s &   $652.5$k & 1.78s\\
\hline
\end{tabular}
\end{table}

Table~\ref{table:parameters} compares the computational efficiency of the proposed models in terms of the parameter count, alongside the inference latency measured on an NVIDIA GeForce RTX 3090 GPU. Across both scenarios, FuSNet demonstrates the lowest latency and the smallest parameter count, whereas LAeNet exhibits the highest latency and the largest number of parameters. Notably, FuSNet and SCoNet benefit from faster training due to their comparatively low latency, although their memory usage increases linearly with the window length. In contrast, LAeNet maintains nearly constant memory usage as the window length increases, owing to its shared LSTM and feedforward network architecture. This design allows efficient parameter utilization for longer windows but results in substantially longer training times due to higher latency.

\section{Application into speech enhancement}

Motivated by the computational efficiency of the proposed architectures, in this section we further evaluate the ReTM estimation accuracy in a speech denoising application \cite{kumar2024speech}.

Denote $\bm{\mathcal{R}}_{AB}^{(N)}$ be the noise sources ReTM. In brief, \cite{kumar2024speech} proposed to enhance the speech from the noisy speech recordings with the known $\bm{\mathcal{R}}_{AB}^{(N)}$, which can be estimated using the noise-only recordings, as 
\begin{align}
\hat{\mathbf{S}} \triangleq \mathbf{M}_A - \bm{\mathcal{R}}_{AB}^{(N)} \mathbf{M}_B 
= \big[ \mathbf{h}_A - \bm{\mathcal{R}}_{AB}^{(N)} \mathbf{h}_B \big] S,
\label{eqn:retm_sss}
\end{align}
where $\mathbf{h}_A$ and $\mathbf{h}_B$ be the acoustic transfer function vectors from the speech source to group $\{A\}$ and $\{B\}$ microphones, respectively. We note that $\mathbf{\hat{S}}$ is a $ Q_A \times 1$ vector consists $Q_A$ copies of estimated target speech signal $S$.

We evaluate speech denoising performance in scenarios \textbf{B} and \textbf{C} (Sec.~\ref {results}), each containing a single speech source, where the ReTM is estimated from 1-minute noise-only recordings. For analysis we use the SDR \cite{vincent2006performance}, and the Short-Time Objective Intelligibility (STOI) \cite{taal2011algorithm}. 

Table~\ref{table:denoising} depicts the speech enhancement accuracy for both \textbf{B} and \textbf{C} scenarios. The results confirm that enhanced performance ($+0.5$ dB in SDR, $13\%$ in STOI) for all methods. Although FuSNet significantly outperforms ReTM estimation on most metrics, its speech denoising performance degrades, confirming the advantage of time-frequency domain ReTM estimation over the time domain approach.
Compared to FuSNet, both LAeNet and SCoNet, demonstrate significantly superior performance. LAeNet achieves the highest performance in both scenarios \textbf{B} and \textbf{C}, with an average SDR of $+10.55$ dB and an average STOI of $37\%$. Whereas, SCoNet ranks second in both scenarios, yielding an average SDR of $+8.41$ dB and an average STOI $33\%$.
In contrast, the traditional method demonstrates comparable performance, although the proposed models, except FuSNet, consistently achieve better results in both scenarios.
From informal listening, we find that FuSNet's denoised speech remains significant echo noise, leading to lower performance, which could be improved through dereverberation algorithms \cite{naylor2010speech,wang2020deep} that we will investigate in future work. The audio samples of this work can be found on GitHub.\footnote{\url{https://github.com/oshanyalegama/Denoised_ReTM_DL}}

\begin{table}[]
\centering
\caption{Evaluation results on the scenarios B \& C.}
\label{table:denoising}
\begin{tabular}{c|cc|cc}
\hline
\multirow{2}{*}{Method} 

  & \multicolumn{2}{c|}{ \textbf{B}} 
  & \multicolumn{2}{c}{ \textbf{C}} \\
\cline{2-5}
  & \multicolumn{1}{c}{SDR} 
  & \multicolumn{1}{c|}{STOI} 
  & \multicolumn{1}{c}{SDR} 
  & \multicolumn{1}{c}{STOI} \\
\hline
Noisy & $-2.70$ & $0.54$ & $-2.70$ & $0.54$ \\
Baseline &  $6.06$ &  $0.87$ & $4.07$ & $0.85$ \\
SCoNet   &  $6.45$ &  $0.87$ &   $4.96$ &  $0.87$ \\
FuSNet  & $2.21$ & $0.80$ & $-1.19$ & $0.67$ \\
LAeNet   & $\bm{8.67}$ &  $\bm{0.92}$ &  $\bm{7.03}$ & $\bm{0.91}$ \\
\hline
\end{tabular}
\end{table}




\section{Conclusion}

We proposed SCoNet, FuSNet, and LAeNet to estimate the ReTM using machine learning. Our proposed approaches use either time or time-frequency domains to model the inter-group spatial relationship. 
Experimental results confirmed that FuSNet consistently outperformed other approaches in terms of accuracy, while SCoNet and LAeNet achieved performance comparable to the baseline method. We further validated the effectiveness of the proposed methods in a speech denoising application. In future work, we aim to extend these models to source separation, speech dereverberation, and investigate optimal microphone grouping strategies for ReTM-based applications.


\section{Acknowledgements}

This work was supported by the Accelerating Higher Education Expansion and Development (AHEAD) operation (Grant No. 6026-LK/8743-LK, World Bank) and a data scholarship from the Linguistic Data Consortium, University of Pennsylvania.
The authors thank Prof. Thushara Abhayapala for fruitful discussions and support on this work.

\section{Use of Generative AI Disclosure}
Generative AI tools were used to assist in drafting and refining portions of the scripts. Additionally, grammar-checking models were utilized to improve language clarity, correctness, and overall readability.

\bibliographystyle{IEEEtran}
\bibliography{ref}

\end{document}